\documentclass[aps,pra,twocolumn,amsmath,amssymb,nofootinbib,superscriptaddress]{revtex4}

\newcommand{\bra}[1]{\langle#1|}
\newcommand{\ket}[1]{|#1\rangle}

\usepackage[colorlinks]{hyperref} %links for citations
\usepackage[pdftex]{graphicx}
\usepackage{mathrsfs}

\begin{document}

\bibliographystyle{apsrev}

\title{Quantum walks with memory -- goldfish, elephants and wise old men}

\author{Peter P. Rohde}
\email[]{dr.rohde@gmail.com}
\homepage{http://www.peterrohde.org}
\affiliation{Centre for Engineered Quantum Systems, Department of Physics and Astronomy, Macquarie University, Sydney NSW 2113, Australia}

\author{Gavin K. Brennen}
\email[]{gavin.brennen@mq.edu.au}
\affiliation{Centre for Engineered Quantum Systems, Department of Physics and Astronomy, Macquarie University, Sydney NSW 2113, Australia}

\author{Alexei Gilchrist}
\email[]{alexei@ectropy.info}
\affiliation{Centre for Engineered Quantum Systems, Department of Physics and Astronomy, Macquarie University, Sydney NSW 2113, Australia}

\date{\today}

\frenchspacing

% Abstract
\begin{abstract}
Quantum walks have emerged as an interesting approach to quantum information processing, exhibiting many unique properties compared to the analogous classical random walk. Here we introduce a model for a discrete-time quantum walk with memory by endowing the walker with multiple recycled coins and using a physical memory function via a history dependent coin flip.  By numerical simulation we observe several phenomena.  First in one dimension, walkers with memory have persistent quantum ballistic speed up over classical walks just as found in previous studies of multi-coined walks with trivial memory function.  However, measurement of the multi-coin state can dramatically shift the mean of the spatial distribution. Second, we consider spatial entanglement in a two-dimensional quantum walk with memory and find that memory destroys entanglement between the spatial dimensions, even when entangling coins are employed. Finally, we explore behaviour in the presence of spatial randomness and find that in contrast to single coined walks, multi-coined walks do not localise and in fact a memory function can speed up the walk relative to a fully decohered multi-coin walker with trivial memory.  We explicitly show how to construct linear optics circuits implementing the walks, and discuss prospects for classical simulation.
\end{abstract}

\maketitle

% Introduction
\section{Introduction}

Quantum computation \cite{bib:NielsenChuang00} is believed to allow certain computational problems to be solved much more quickly than on classical computers. The standard approach to quantum computation is using the circuit model, whereby an algorithm is decomposed into a set of quantum gates. Another contender is the measurement based model \cite{bib:Raussendorf01, bib:Raussendorf03}, whereby a highly entangled state is prepared and computation proceeds via only single qubit measurements. More recently, the \emph{quantum walk} formalism, the quantum analogy of a classical random walk, has emerged as an interesting approach to implementing quantum computational tasks  \cite{bib:ADZ, bib:AAKV, bib:Kempe08, bib:Salvador12}. Here a \emph{walker}, i.e. a particle, is located at vertices in a graph and is allowed to coherently `hop' along the edges of the graph to other vertices. This approach has proved fruitful for algorithm design \cite{bib:ChildsCleve03, bib:ShenviKempe03, bib:Ambainis04, bib:ChildsGoldstone04, bib:Magniez07, bib:Santha08, bib:Tulsi08, bib:Berry11a, bib:Berry11} and is known to be universal for quantum computation \cite{bib:Childs09, bib:Lovett10}. Numerous experiments have begun to demonstrate elementary quantum walks, particularly photonic implementations where the walker is a single photon \cite{bib:Hagai08, bib:Schreiber10, bib:Broome10, bib:Peruzzo10, bib:Schreiber11b, bib:Matthews11, bib:Schreiber12, Sansoni12}. The standard models consider single walkers. However steps have been made towards implementing multi-walker quantum walks \cite{bib:RohdeSchreiber11, bib:RohdeSchreiber12, bib:Peruzzo10}.

In the usual discrete-time quantum walk model a walker has a \emph{coin} value associated with it, which specifies the direction the walker will take when propagating through the graph. In this paper we consider the case where multiple coins are employed, which can be interpreted as memory of previous coin values. We show that this memory drastically affects the evolution of the quantum walk, and indeed can result in a transition from quantum statistics to classical statistics. We study the effects of history-dependent coins and find that this can lead to various diffusive phenomena. We consider the effect that measuring memory elements has on the evolution of the walk, and discuss the rate of spread of the probability distribution as a function of the memory length. We explicitly show how to experimentally construct an optical quantum walk with memory and discuss the challenges of classical simulation of such systems.

Numerous authors have considered memory effects in the context of classical random walks \cite{bib:Dickman03, bib:Hod04, bib:Schutz04, bib:Keshet05, bib:Paraan06, bib:daSilva06, bib:Cressoni07, bib:Kenkre07, bib:Harris09}, and some steps have also been made in the quantum context \cite{bib:Brun03, bib:Flitney04, bib:McGettrick10}, which we build upon.

% Quantum walk formalism for linear graphs
\section{Quantum walk formalism for linear graphs}

We begin by defining a walker as a bipartite system, $\ket{x,c}$, where $x$ is the \emph{position} of the walker in a graph and $c$ is the \emph{coin} value which dictates the direction of the walker. In the standard discrete-time quantum walk formalism for a single walker on a linear graph, the evolution is decomposed into two steps, \emph{coin} ($C$) and \emph{step} ($S$), defined as,
\begin{eqnarray}
C: && \ket{x,c} \to \sum_j A_{c,j}^{(x)} \ket{x,j}, \nonumber \\
S: && \ket{x,c} \to \ket{x+c,c},
\end{eqnarray}
where $A^{(x)}$ is a unitary coin matrix defining the transition amplitudes at position $x$. The coin value takes values $\pm 1$ (right or left respectively). The coin operator coherently manipulates the coin value, putting it into a superposition of left and right, while the step operator updates the position value according to the newly chosen coin value. After the coin operator, the coin value can be interpreted as the direction the walker will propagate at the the next step, whereas after the step operator it can be regarded as memory of the previous coin direction. After evolving $t$ time steps the output state is \mbox{$\ket{\psi_\mathrm{out}}=(SC)^t\ket{\psi_\mathrm{in}}$}. This formalism could be logically extended to operate on graphs of higher order by extending the coin space beyond $\pm 1$, but we will initially focus on the case of linear graphs for ease of exposition.

The rate of spread of a quantum walk is quantified by the variance:  
\begin{equation}
\sigma^2(t) = \sum_i p_i(t)(i-\mu)^2.
\end{equation}
where \mbox{$\mu=\sum_i p_i i$}, and $p_i(t)$ is the probability to measure the walker at position $i$ after a number of time steps $t$.
A classical random walk on a line has variance \mbox{$\sigma^2(t)=t$} whereas the quantum walk on a line \mbox{$\sigma^2(t)\propto t^2$} where the proportionality constant depends on the initial state and the choice of coin rotation \cite{bib:Kempe08}.  In the presence of decoherence, either acting on the coin or spatial degrees of freedom, the quantum walk eventually behaves like a classical random walk with linear dispersion \cite{bib:Kendon}.

% Formalism with memory
\section{Formalism with memory}

\begin{figure}
\includegraphics[width=\columnwidth]{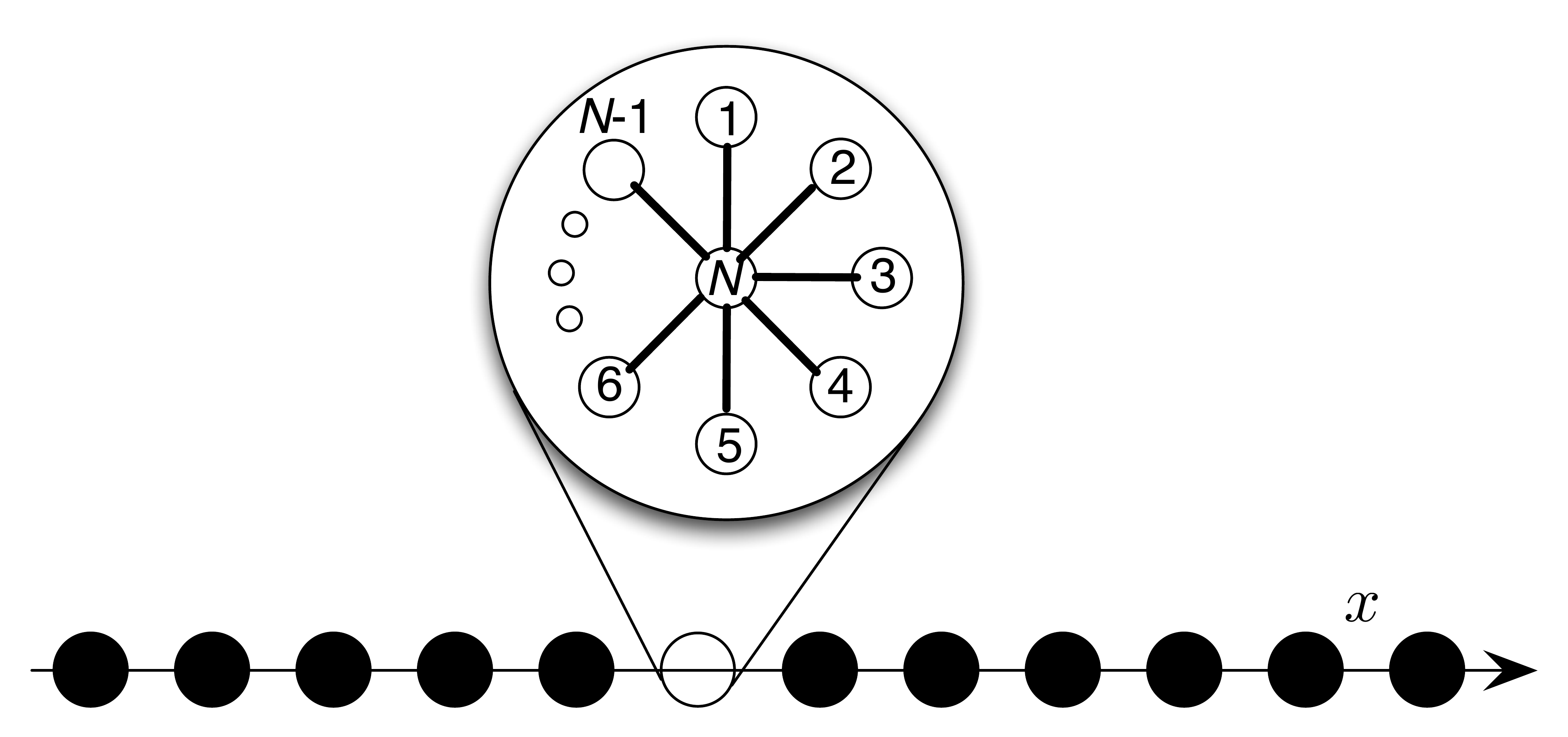}
\caption{Schematic of the quantum walk with memory.  The walker can be viewed as a molecule carrying $N$ spin-$1/2$ particles.  The coin flip operation on spin $N$ is a unitary rotation with angle that is determined by the total polarisation of the other \mbox{$N-1$} memory spins (i.e. a memory function).  As described in Sec. \ref{physmodel} the unitary can be generated by an Ising interaction with the indicated coupling graph.  The centre spin $N$ then performs the conditional shift operation on the entire molecule in space.  Finally, in the memory update step, the spins are cyclically permuted. } \label{WalkerFig}
\end{figure}

In the case of an arbitrary number of memory elements we introduce additional memory Hilbert spaces. We focus on discrete-time quantum walks since no coins are present in the continuous-time quantum walk, and therefore memory does not naturally arise. With $N$ memory elements the walker is defined as \mbox{$\ket{x,c_1,\dots,c_N}$}, and  we decompose the evolution into three stages by adding an additional \emph{memory update} ($M$) operator. Then the evolution can be decomposed into,
\begin{eqnarray} \label{eq:ket_rules}
C: && \ket{x,c_1,\dots,c_N} \to \sum_j A_{c_N,j}^{(x)} \ket{x,c_1,\dots,c_{N-1},j}, \nonumber \\
S: && \ket{x,c_1,\dots,c_N} \to \ket{x+c_N,c_1,\dots,c_N}, \nonumber \\
M: && \ket{x,c_1,\dots,c_N} \to \ket{x,c_N,c_1,\dots,c_{N-1}},
\end{eqnarray}
where $c_i$ is the coin value at memory element $i$, i.e. memory of the coin value $i$ steps ago.  All the operations can be done quasi-locally in space (see Fig. \ref{WalkerFig}). The coin operator coherently manipulates the last memory element into a superposition of left and right, the step operator updates the current position according to the last memory element, and the memory update operator cyclically permutes the memory elements such that \mbox{$c_i\to c_{i+1}$} (i.e. what was the $i$th memory element next becomes the \mbox{$(i+1)$th} memory element).  
When $N=1$ we refer to this as a \emph{goldfish walk} (goldfish have poor short-term memory), and when $N=t$ we refer to this as an \emph{elephant walk} (elephants have exceptional long-term memory).

Our definition of a walk with memory is similar to that by Brun \emph{et al.} \cite{bib:Brun03} who introduced the use of cyclic permutations to update the memory registers but without a memory dependent coin flip.  Flitney \emph{et al.} \cite{bib:Flitney04}, considered the case of some form of memory dependent coin flip in the context of Parrondo games.   See also McGettrick \emph{et al.} \cite{bib:McGettrick10} for a treatment of quantum walks with a length-two memory. 
% Brun \emph{et al.} especially focus on the moments of the probability distribution with different memory lengths and compare this to the classical case. 
One of the key results of that work is that quantum walks with recycled coins do not decohere provided that the coins are not recycled faster than every other time step.  If, however, the number of coins does equal the number of time steps then the walks converges to the classical random walk behaviour.   Here we give the first systematic numerical study of the effect of memory dependent coin flips on the behaviour of quantum walks. 

One might ask why we have chosen the coin operator to manipulate the last memory element. This is chosen to enforce the interpretation that $c_i$ represents memory of the coin $i$ steps ago. If, say, the first memory element were updated, this interpretation would no longer apply.

After evolving $t$ time steps the output state is \mbox{$\ket{\psi_\mathrm{out}}=(MSC)^t\ket{\psi_\mathrm{in}}$}. To enforce elegant symmetry into the output state henceforth we symmetrize the input state to be of the form,
\begin{equation}
\ket{\psi_\mathrm{in}} =\ket{0}\otimes (\ket{-1,\dots,-1}+\ket{+1,\dots,+1})/\sqrt{2},
\end{equation}
and additionally employ a balanced coin,
\begin{equation} \label{eq:P_coin}
A^{(x)} =e^{-i\frac{\pi}{4}\sigma^x}= \frac{1}{\sqrt{2}} \left( \begin{array}{cc}
1 & -i \\
-i & 1 \\
\end{array} \right) \,\, \forall \,x,
\end{equation}
which applies equal transition amplitudes in the left and right directions at each position.  This particular form of highly entangled input for the coin state is convenient for visualisation purposes but is not necessary. 
 In fact a product state input of the form \mbox{$\ket{0}\otimes \ket{-1}^{\otimes N}$} produces qualitatively similar behaviour with respect to the variance. 

\begin{figure}[!htb]
\includegraphics[width=0.6\columnwidth]{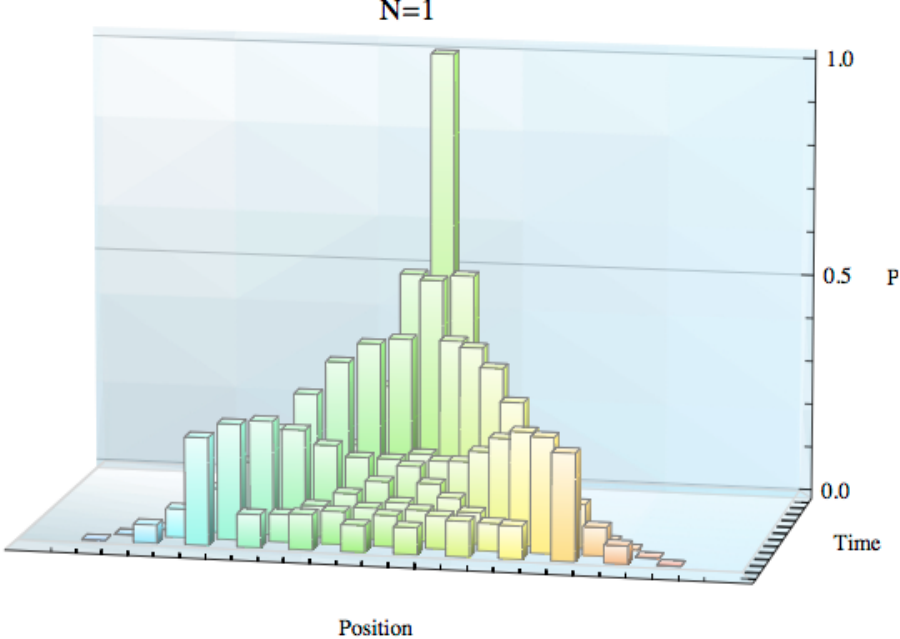} \\
\includegraphics[width=0.6\columnwidth]{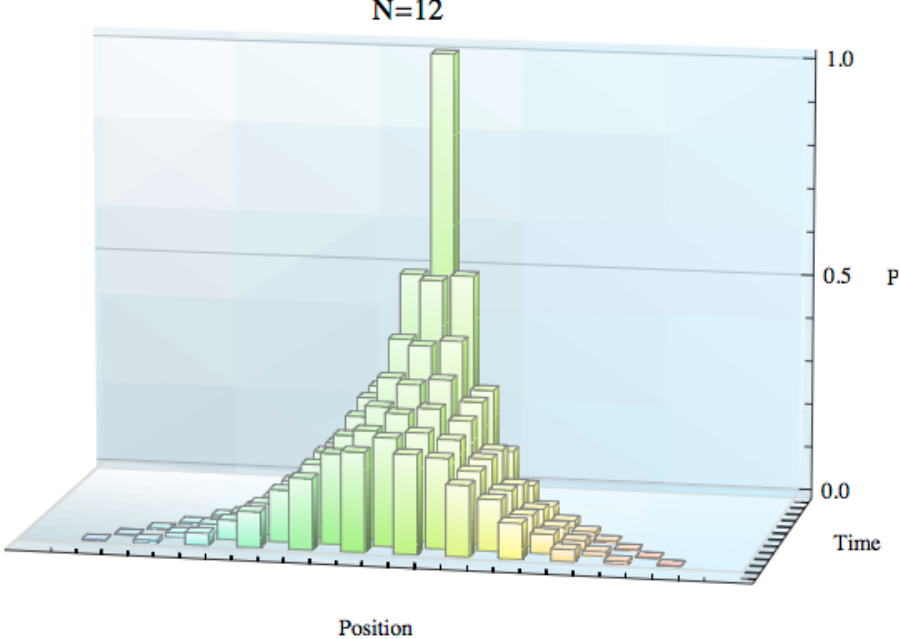}
\caption{(Colour online) Time evolution over 12 steps in the cases of a goldfish walk ($N=1$), where we observe the usual double-peaked quantum walk ballistic spreading (top) and an elephant walk ($N=t=12$), where we observe classical random walk statistics (bottom). Note that even though classical statistics are observed in the elephant walk, no decoherence has taken place and the state is still pure.  This is the limiting case of the Brun \emph{et al.} result for walks with an equal number of coins and walk steps.} \label{fig:plots_3d}
\end{figure}

\begin{figure*}[!htb]
\includegraphics[width=0.5\columnwidth]{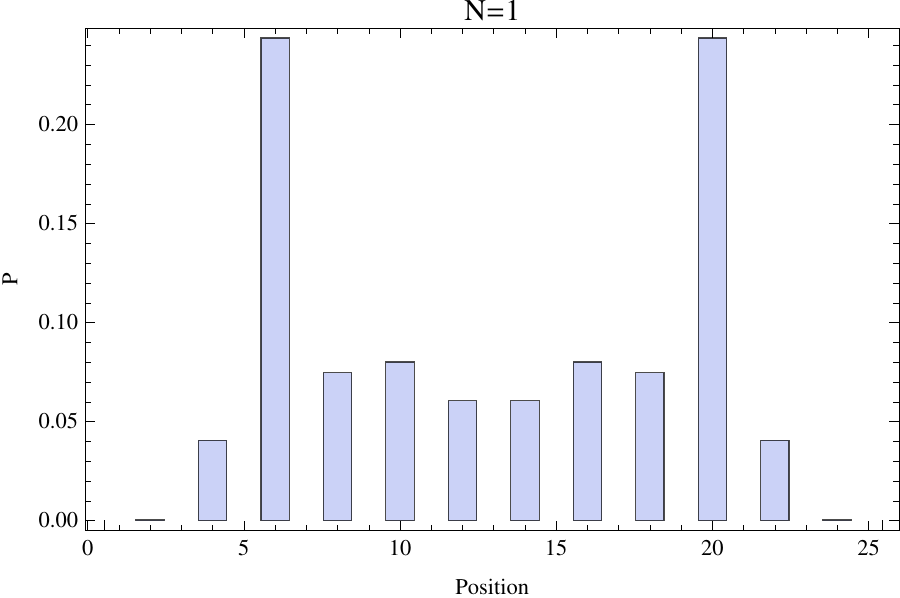}
\includegraphics[width=0.5\columnwidth]{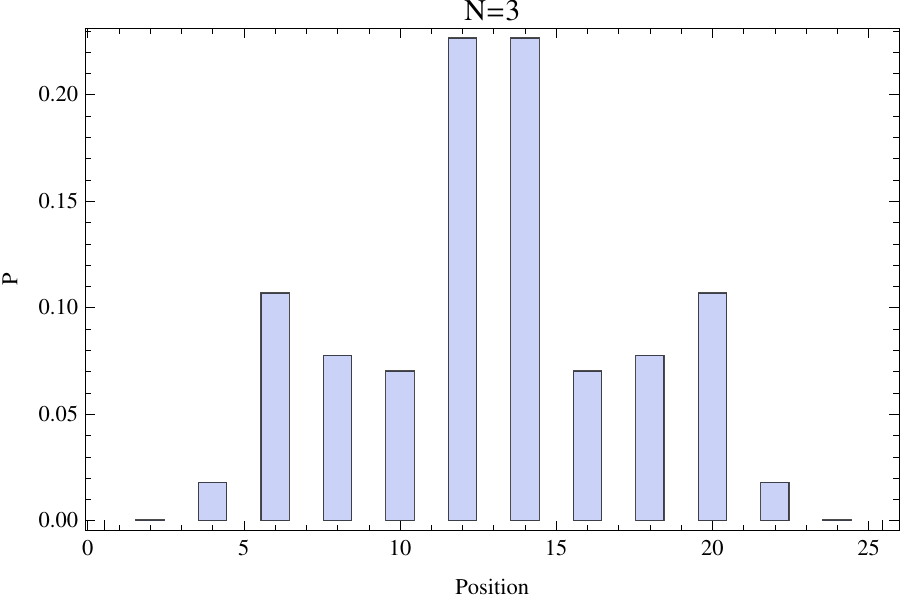}
\includegraphics[width=0.5\columnwidth]{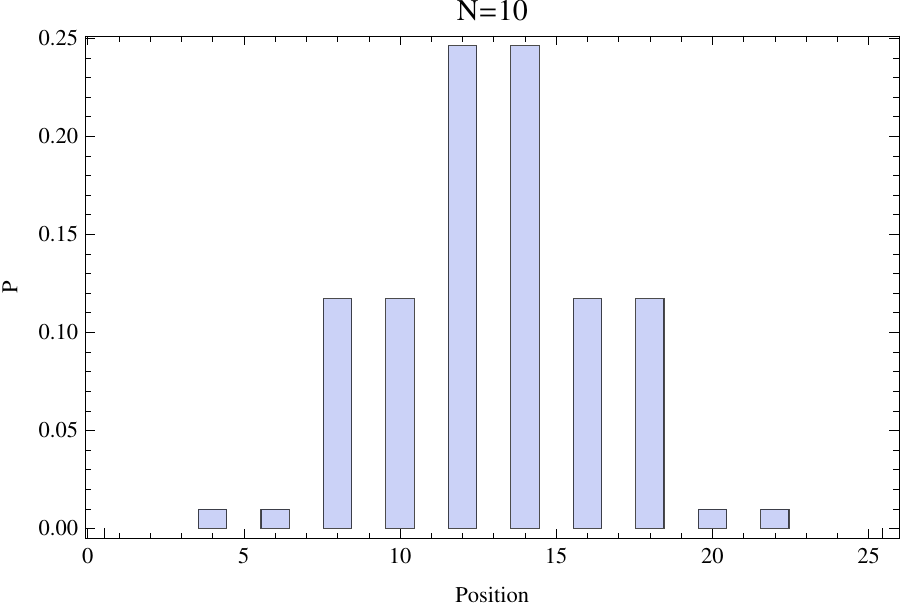}
\includegraphics[width=0.5\columnwidth]{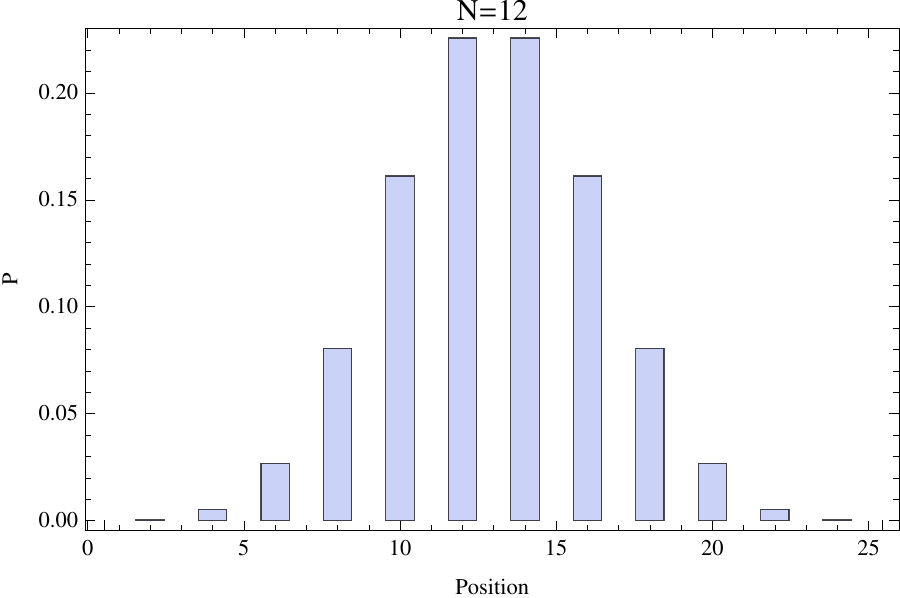}
\caption{(Colour online) Output distributions for \mbox{$t=12$}, with different memory lengths $N$. From left to right: \mbox{$N=1$} (goldfish walk), \mbox{$N=3$}, \mbox{$N=10$} and \mbox{$N=t=12$} (elephant walk). In the case of a goldfish walk we observe the usual double-peaked distribution of a standard single-coined quantum walk, and in the case of an elephant walk we observe the classical statistics of a binomial distribution.  } \label{fig:outputs_memory}
\end{figure*}

The full time evolution over 12 steps of the limiting goldfish and elephant walks are shown in Fig. \ref{fig:plots_3d}. The final output position probability distributions (summed over all coin values) for different memory sizes are illustrated in Fig. \ref{fig:outputs_memory}. In the case of a goldfish walk we observe the usual double-peaked distribution of a standard quantum walk with a single coin, and in the elephant walk we observe the classical statistics of a binomial distribution.

The reason we observe classical statistics for the elephant walk is that the walker always has full memory of its entire evolution. Thus, no two trajectories interfere and therefore evolve independently of one another, giving rise to a classical distribution. An alternate way of thinking about this is that for \mbox{$t<N$} we are effectively using a fresh coin at each step. Thus, we observe classical behaviour for \mbox{$t<N$}, after which quantum behaviour emerges. In general there is a high level of entanglement between the position and memory subsystems. Note the elephant walk yields the same distribution one would obtain for a goldfish walk where the coin value is measured at each time step, or equivalently the coin value undergoes decoherence at each time step, as was demonstrated experimentally by Broome \emph{et al.} \cite{bib:Broome10}. Importantly, even though we observe classical statistics in the elephant walk, our state is nonetheless a highly entangled pure state and no decoherence has taken place. The evolution is unitary and by reversing time we can always evolve back to the input state, which is not possible in the truly classical case.

% Coin measurement & decoherence
\section{Coin measurement \& decoherenece}

\begin{figure*}[!htb]
\includegraphics[width=0.5\columnwidth]{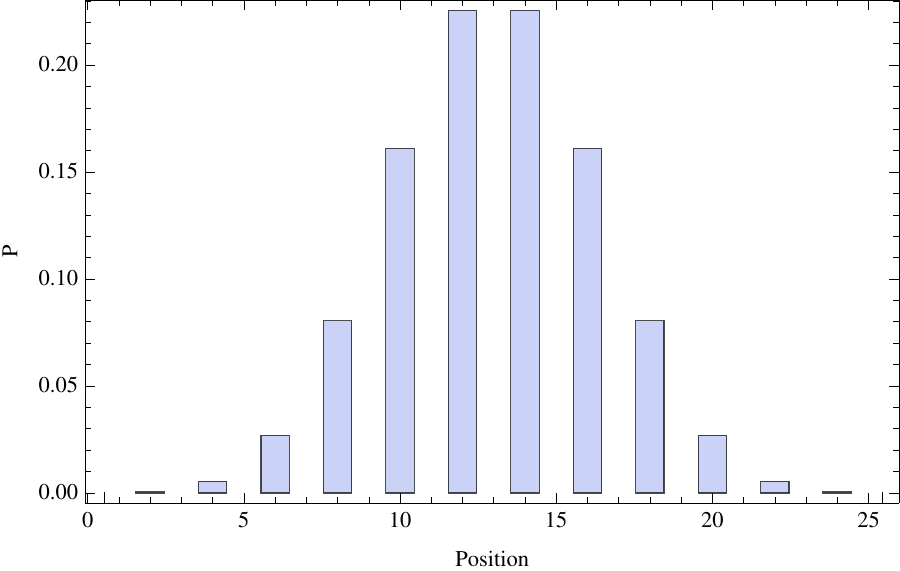}
\includegraphics[width=0.5\columnwidth]{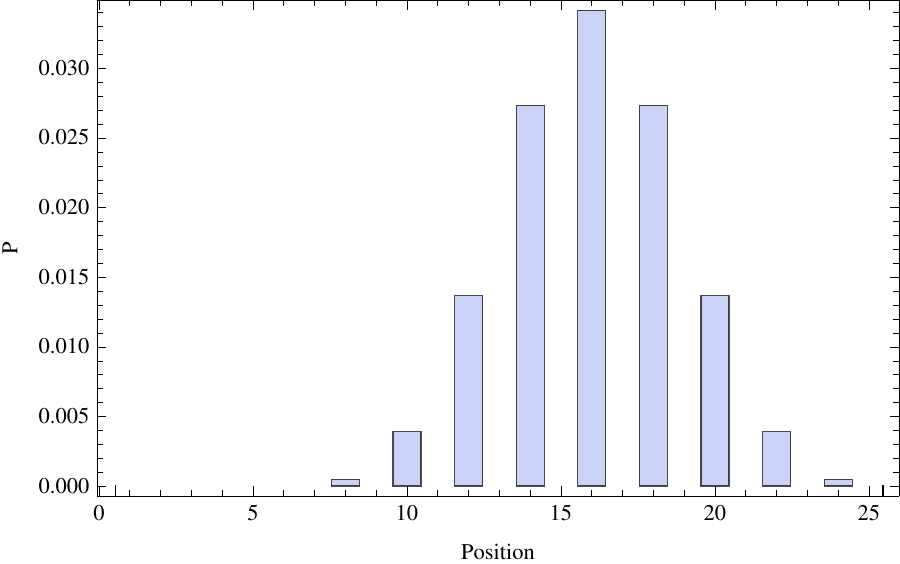}
\includegraphics[width=0.5\columnwidth]{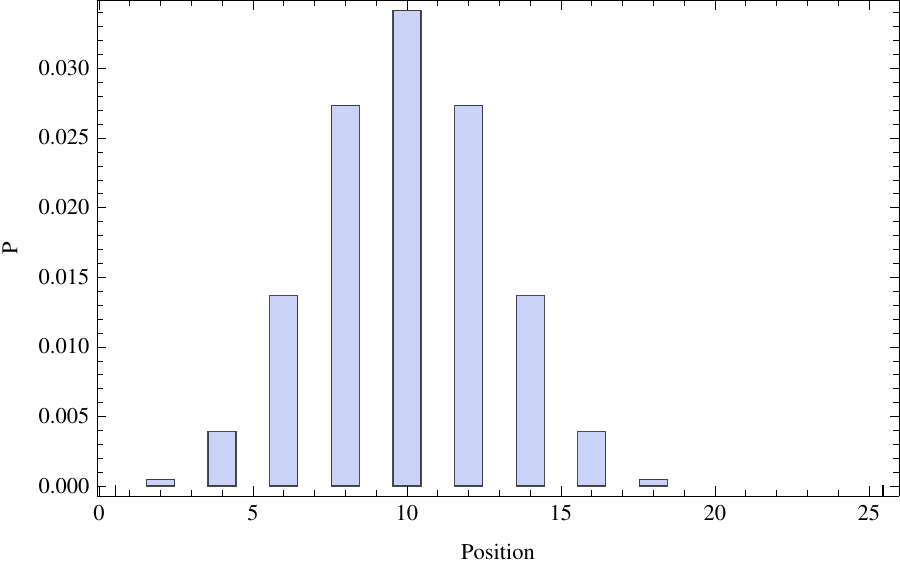}
\caption{(Colour online) Output distributions for an elephant walk ($N=t=12$). (left) no measurement, (middle) measurement of the last four coins with outcomes \mbox{$(1,1,1,1)$}, (right) measurement of the last four coins with outcomes \mbox{$(-1,-1,-1,-1)$}.} \label{fig:outputs_postselect}
\end{figure*}

We now consider the evolution of the quantum walk when we measure some of the coins. In Fig. \ref{fig:outputs_postselect} we plot the output distribution for a an elephant walk with and without measurement of the last four coins in the memory. After measurement we are left with a usual quantum walk distribution, which has been shifted according to the measurement outcomes of the memory registers and allowed to run for a shorter period of time. Specifically, measuring the last $k$ memory registers, with outcomes $o_i$ `resets' the quantum walk and shifts its evolution to have an origin at \mbox{$x_0=\sum_{i=1}^k o_i$} and allowed to run for \mbox{$t-k$} steps. This behaviour would ordinarily be observed in a standard quantum walk if measurements were performed intermittently during the evolution. But here the measurement can be performed after the evolution, which projects the distribution into a region constrained by the measurement outcomes. In the case of an elephant walk, if we measure all the coins we can effectively reconstruct the entire path the walker followed to reach its destination.

If decoherence is present in the memory registers, this is equivalent to the environment measuring the respective registers without giving us information about the measurement outcome. Thus, decoherence in the memory will yield a mixed state over the distributions obtained by different memory measurement outcomes. This can lead to coherence over different length scales. Suppose the environment measures the last coins and is able to distinguish between \mbox{$(1,\dots,1)$} and \mbox{$(-1,\dots,-1)$}, which we acknowledge is a little artificial. This will yield a mixture of two coherent distributions located at different origins on the line, as per Fig. \ref{fig:outputs_postselect}. Each sub-distribution exhibits full coherence, whereas the total distribution is a highly mixed state. If the two sub-distributions are non-overlapping, tracing out the positions associated with one distribution will yield a pure state over the remaining positions.

See Refs. \cite{bib:BrunCarteret03, bib:BrunHillary03} for a discussion on quantum walks with decoherent coins, and Broome \emph{et al.} \cite{bib:Broome10} for an experimental photonic implementation of a quantum walk with decoherent coins showing the transition from quantum to classical as the decoherence is increased.

% In Fig. \ref{fig:spread} we plot the variance of the probability distribution, $$, $\mu=\sum_i p_i i$, for a $t=12$ quantum walk with varying memory lengths $N$. In the case of the elephant walk we previously observed classical statistics. Thus we see a linear rate of increase in the variance against time. In the other limiting case of the goldfish walk we reduce to a standard quantum walk and observe the quadratically faster rate of spread. In between there is a clear transition between the two.

%\begin{figure}[!htb]
%\includegraphics[width=0.9\columnwidth]{spread}
%\caption{(Colour online) Rate of spread of the variance of the position probability distribution against time for different memory lengths. For the goldfish walk ($N=1$) we observe the usual quadratic rate of spread for a normal coined quantum walk, whereas for an elephant walk ($N=t=12$) we observe the linear rate of spread associated with a classical random walk.} \label{fig:spread}
%\end{figure}

% Two spatial dimensions
\section{Two spatial dimensions}

\begin{figure*}[!htb]
\includegraphics[width=0.5\columnwidth]{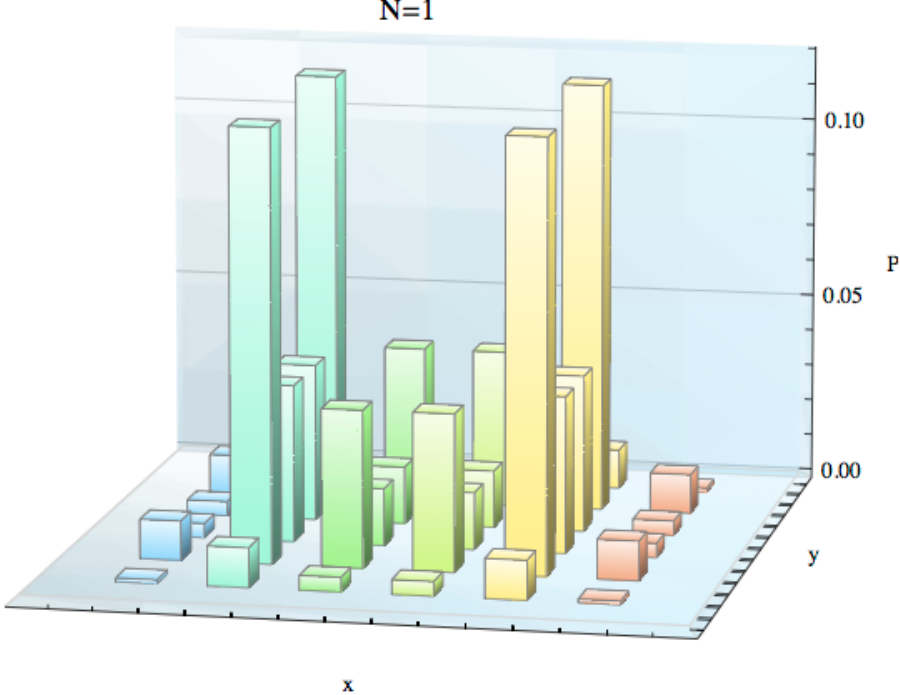}
\includegraphics[width=0.5\columnwidth]{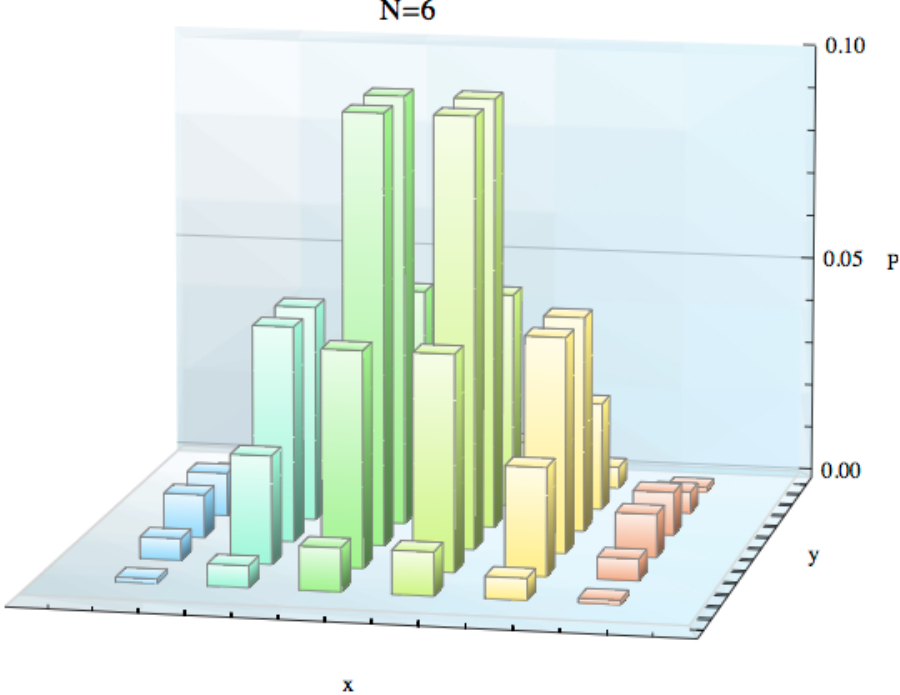}
\includegraphics[width=0.5\columnwidth]{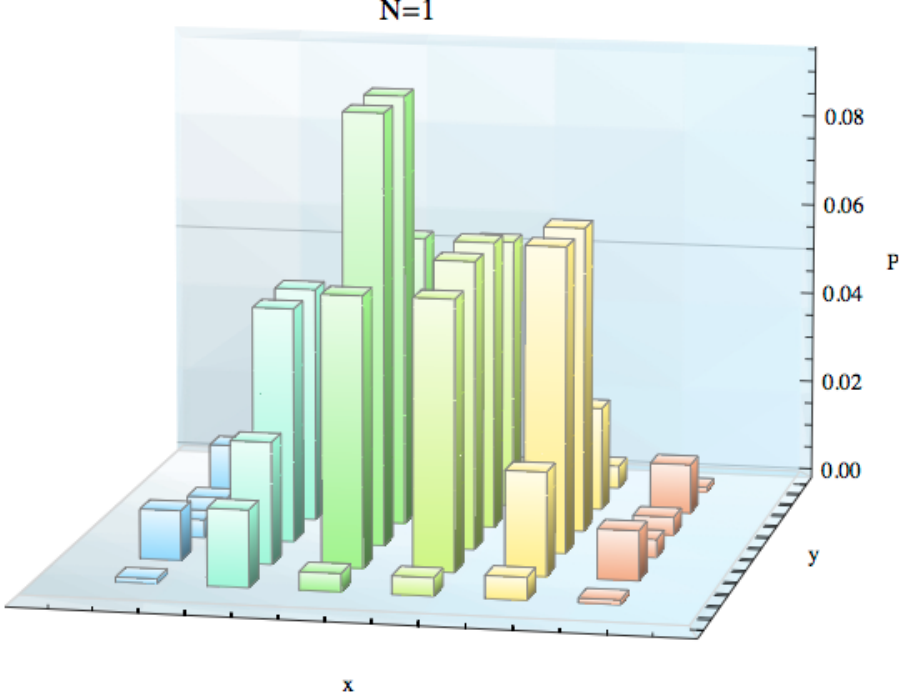}
\includegraphics[width=0.5\columnwidth]{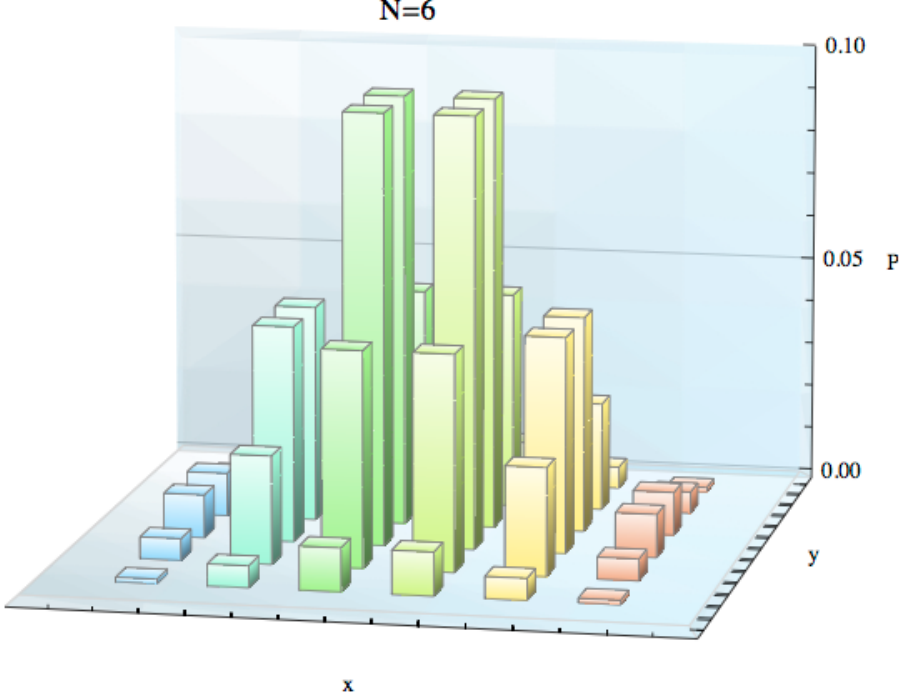}
\caption{(Colour online) Spatial distribution of a quantum walk on a 2D lattice with $t=6$ using separable (Eq. \ref{2Dsep}) and entangling (Eq. \ref{2Dent}) coin flips: (1) Goldfish walk, separable coins; (2) Elephant walk, separable coins; (3) Goldfish walk, entangling coins; (4) Elephant walk, entangling coins. The distribution is separable across the two spatial degrees of freedom when either separable coins are employed or there is full memory. The only instance where the distribution is non-separable (i.e. where there is entanglement between the two spatial degrees of freedom) is when entangling coins are employed and there is no memory.} \label{fig:two_dimensions}
\end{figure*}

Next we turn our attention to a quantum walk on a two dimensional lattice. This is easily constructed by taking the formalism introduced earlier and expanding the coin degree of freedom to have four basis states (up, down, left and right). Now the basis states are of the form \mbox{$\ket{x,y,c_1,\dots,c_N}$}, where $x$ and $y$ denote the two spatial dimensions. We begin with the walker localised at the centre of the graph and consider the cases where the coin is separable
\begin{equation}
A_{\rm 2D}^{(x)}=e^{-i\frac{\pi}{4}\sigma^x}\otimes e^{-i\frac{\pi}{4}\sigma^x} \,\, \forall \, x
\label{2Dsep}
\end{equation}
where \mbox{$e^{-i\frac{\pi}{4}\sigma^x}$} is the coin from Eq. \ref{eq:P_coin}  (i.e. the coin acts on the two spatial dimensions independently) or maximally entangling,
\begin{equation}
A_{\rm 2D\ ent}^{(x)}=(e^{-i\frac{\pi}{4}\sigma^x}\otimes e^{-i\frac{\pi}{4}\sigma^x})\cdot \left( \begin{array}{cccc} 1 & 0 & 0 & 0 \\ 0 & 1 & 0 & 0 \\ 0 & 0 & 0 & 1 \\ 0 & 0 & 1& 0 \end{array} \right) \,\, \forall \, x,
\label{2Dent}
\end{equation}
both with and without memory. The final probability distributions are illustrated in Fig. \ref{fig:two_dimensions}. It can be visually seen that all the distributions are separable across the two spatial degrees of freedom, except in the case where entangling coins are employed \emph{and} there is no memory.

Interestingly, even when entangling coins are employed, we do not observe spatial entanglement when memory is introduced. Thus, not only does memory reduce the walk to a classical binomial distribution, but it also destroys entanglement between the two spatial degrees of freedom. When the probability distribution is separable this implies the walker's $x$ value is independent of its $y$ value and we will always observe the same marginal $x$ distribution. Thus, measurement along one direction has no impact on the measured distribution in the other. On the other hand, with spatial entanglement (i.e. inseparability), measurement in the $x$ direction projects the $y$ distribution onto a function of the measured $x$ outcome.

\begin{figure}[!htb]
\includegraphics[width=\columnwidth]{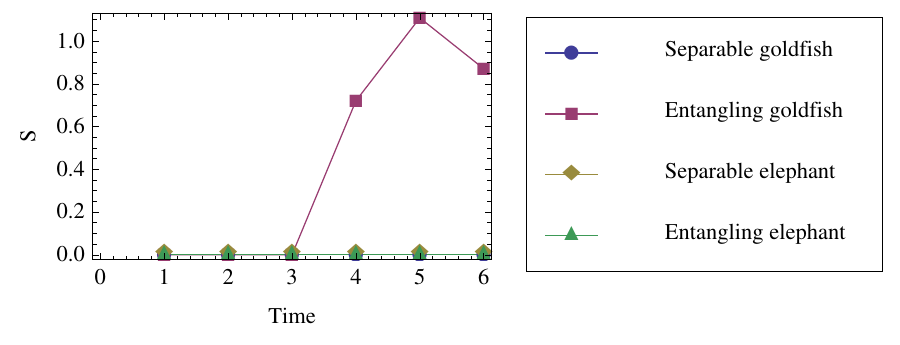}
\caption{(Colour online) Spatial entanglement dynamics of 2D goldfish and elephant quantum walks, with separable and entangling coins. We only observe spatial entanglement when the coin is entangling and there is no memory.} \label{fig:ent_dyn}
\end{figure}

To quantify the spatial entanglement, in Fig. \ref{fig:ent_dyn} we plot the spatial entanglement dynamics against time, confirming that spatial entanglement only persists with entangling coins and no memory. The entanglement metric used is that described in Ref. \cite{bib:RohdeSchreiber11}. We diagonalise the spatial probability distribution matrix and then calculate the Shannon entropy of the diagonal elements, \mbox{$S=-\sum_i \lambda_i\,\mathrm{log}_2\, \lambda_i$}. When the spatial distribution is separable (i.e. there is no spatial entanglement), there is only one diagonal element and \mbox{$S=0$}. For non-separable distributions we have \mbox{$S>0$}, a signature of entanglement.

% History-dependent coins
\section{History-dependent coins}

Until now we have considered coins that are independent of the full memory history and only depend on the last memory element. We now consider the scenario where the coin operator is a function of the full previous history. We refer to this as a \emph{wise old man walk} (wise old men have varying degrees of memory -- some good, some horrendous -- but they always base their actions on their past).

Specifically, we will assume biased coins of the form,
\begin{equation} \label{eq:hist_dep_coin}
A^{(x)}(\theta) = e^{i\theta\sigma^y} =\left( \begin{array}{cc}
\mathrm{cos}\,\theta & \mathrm{sin}\,\theta \\
-\mathrm{sin}\,\theta & \mathrm{cos}\,\theta \\
\end{array} \right) \,\, \forall \,x,
\end{equation}
where the angle $\theta$ is computed from the memory history by
\begin{equation} \label{eq:theta}
\theta = \frac{\pi}{4} + \phi \frac{\pi}{4} \frac{\sum_{i=1}^{N-1}c_i}{N-1},
\end{equation}
and \mbox{$0\leq \phi \leq 1$} is a parameter which specifies to what extent the memory is taken into account when choosing the coin operator. We refer to $\theta$ as the \emph{memory function}, and it determines the diffusive properties of the walk. Of course this definition only works when \mbox{$N\geq 2$}. When $\phi=0$ the coin operator reduces to a balanced Hadamard coin, independent of the history, and we reduce to walks similar to those previously discussed.

\begin{figure*}[!htb]
\includegraphics[width=0.65\columnwidth]{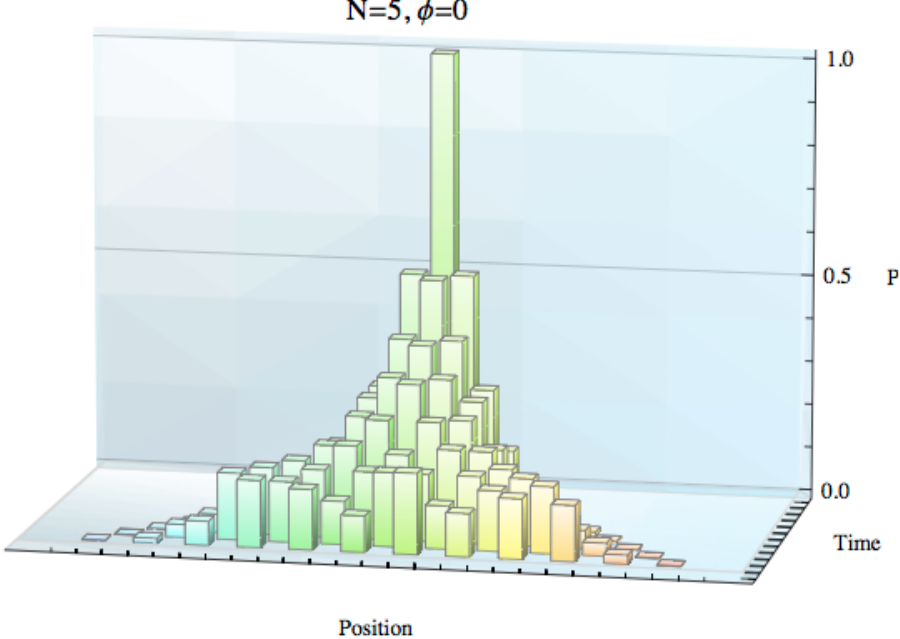}
\includegraphics[width=0.65\columnwidth]{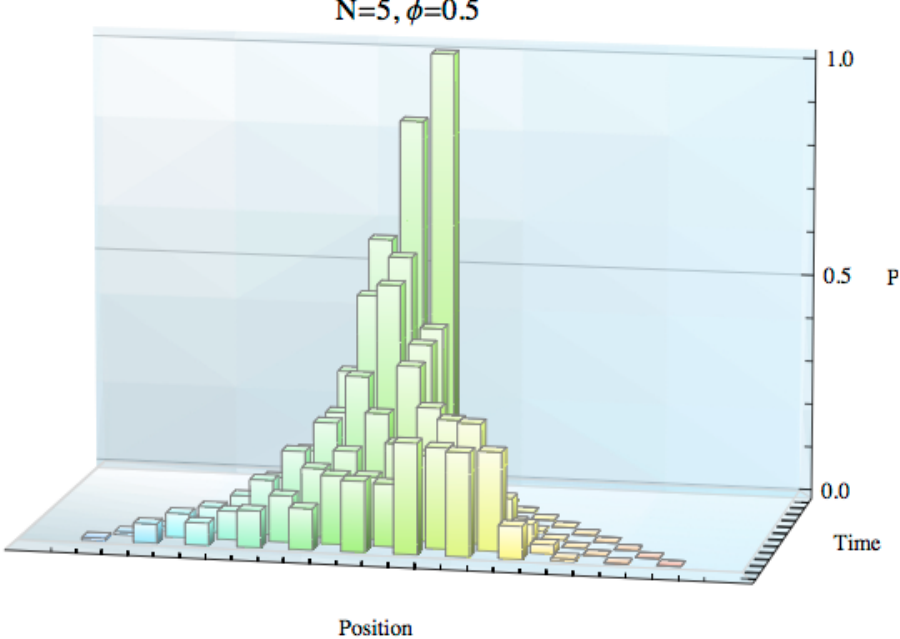}
\includegraphics[width=0.65\columnwidth]{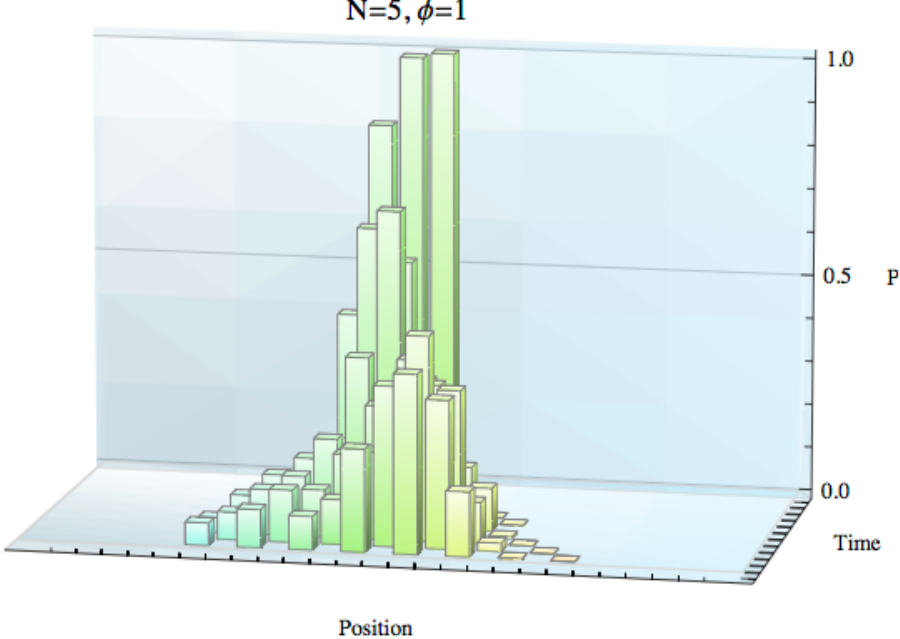}
\caption{(Colour online) Time evolution over 12 steps for a wise old man walk with $N=5$ and memory function from Eq. \ref{eq:theta}. (left) $\phi=0$ (a Hadamard walk), (middle) $\phi=0.5$ and (right) $\phi=1$. As $\phi$ increases, diffusion in one direction becomes more limited than the other, yielding an asymmetric distribution.} \label{fig:wise_phi}
\end{figure*}

\begin{figure}
\includegraphics[width=\columnwidth]{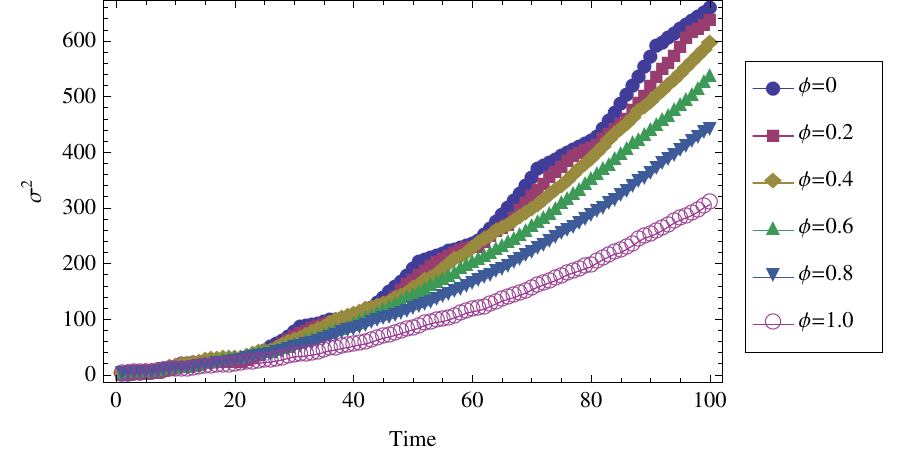}
\caption{(Colour online) Evolution of the variance of the probability distribution against time for an $N=5$ wise old man walk for different history parameters $\phi$ and memory function from Eq. \ref{eq:theta}. As $\phi$ increases, diffusion decreases. All curves fit closely to quadratic distributions.} \label{fig:wise_spread}
\end{figure}

In Fig. \ref{fig:wise_phi} we plot the time evolution of a wise old man walk with $N=5$ for different history parameters $\phi$ and memory function from Eq. \ref{eq:theta}. This time we use the input state \mbox{$\ket{\psi_\mathrm{in}}=\ket{0,1,\dots,1}$}, i.e. we do not symmetrize the input state . When $\phi=0$ we reduce to a normal Hadamard walk. As $\phi$ increases we observe reduced dispersion, biased in one direction, yielding an asymmetric position distribution. To quantify this, in Fig. \ref{fig:wise_spread} we plot the variance of the distribution against time for different values of $\phi$, confirming that higher $\phi$ results in reduced dispersion. All distributions in Fig. \ref{fig:wise_spread} closely fit to quadratic curves.

Fig. \ref{fig:wise_phi} can be understood as follows. Assume $\phi=1$. If the walker has a long history of moving to the right, $\theta\to \pi/2$, and the coin operator reduces to the bit-flip matrix, \mbox{$A^{(x)}(\frac{\pi}{2})=X=\left( \begin{array}{cc} 0 & 1 \\ 1 & 0 \\ \end{array} \right) \,\, \forall \, x$}. Then the walker favours switching direction and heading back towards the origin. On the other hand, if the walker has a long history of moving to the left, $\theta\to 0$, the coin operator reduces to the identity matrix, \mbox{$A^{(x)}(0)=I=\mathrm{diag}(1,1) \,\, \forall \, x$}, and the walker will favour continuing its present course. Thus in one direction the walker resists further diffusion, whilst in the other it does not, giving rise to the asymmetric distribution seen in Fig. \ref{fig:wise_phi}. The latter property is better understood by considering the input state \mbox{$\ket{\psi_\mathrm{in}}=\ket{0,-1,\dots,-1}$}. In this case the coin is the identity matrix, and remains so, and the walker linearly shoots to the left without entering a superposition (graphic not shown). For this reason in the Fig. \ref{fig:wise_phi} we have only considered the \mbox{$\ket{\psi_\mathrm{in}}=\ket{0,1,\dots,1}$} term, which nicely illustrates the resistive diffusive behaviour in the rightward direction.

\begin{figure}
\includegraphics[width=0.6\columnwidth]{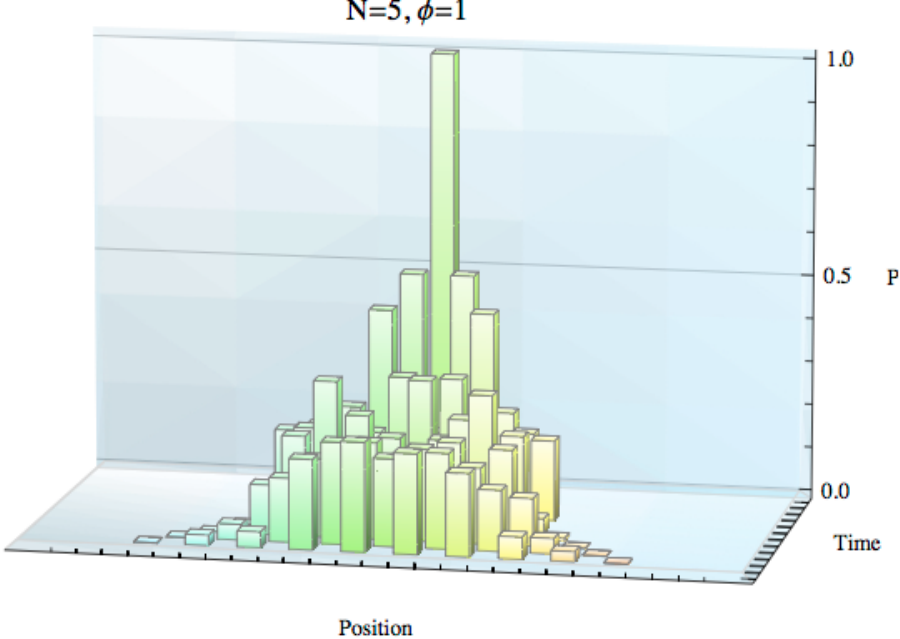}
\caption{Time evolution over 12 steps for a wise old man walk with $N=5$ and memory function from Eq. \ref{eq:second_diff}. Diffusion is restricted in both directions, yielding a much more localised distribution than Fig. \ref{fig:wise_phi}(left).} \label{fig:second_diff}
\end{figure}

Of course there is nothing unique about our choice of memory function in Eq. \ref{eq:theta}. Any function could be chosen, giving rise to a plethora of different diffusive phenomena. Another simple example is to choose the memory function to be,
\begin{equation} \label{eq:second_diff}
\theta = \frac{\pi}{4} + \phi \frac{\pi}{4} \frac{|\sum_{i=1}^{N-1}c_i|}{N-1}.
\end{equation}
In this instance the coin reduces to a bit-flip only when the walker's history is consistently in the same direction and to a Hadamard when the history is balanced. Thus, this memory function will resist excessive diffusion in both directions, yielding the much more localised distribution shown in Fig. \ref{fig:second_diff} (once again with a symmetrized input state).

\begin{figure}
\includegraphics[width=0.6\columnwidth]{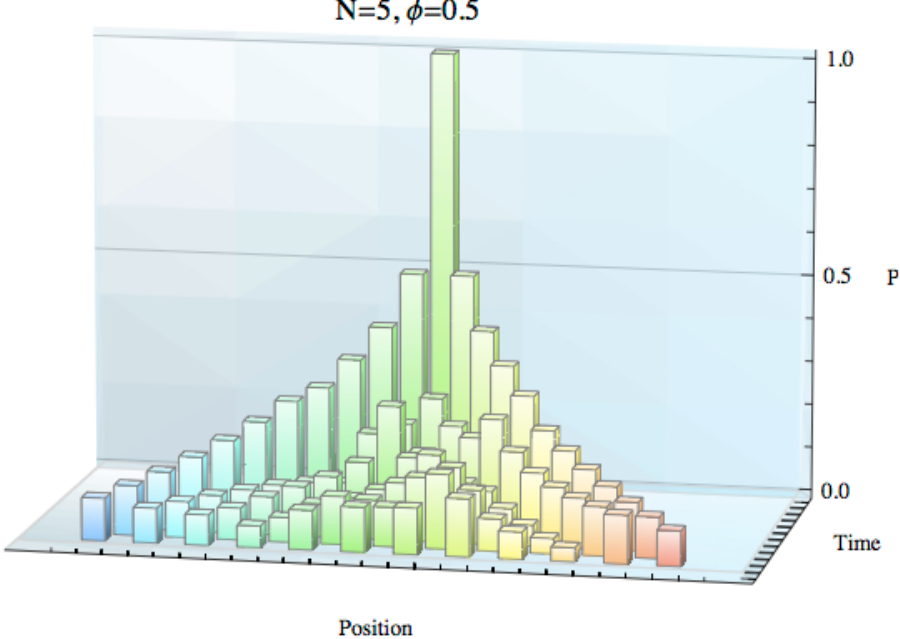}
\includegraphics[width=0.6\columnwidth]{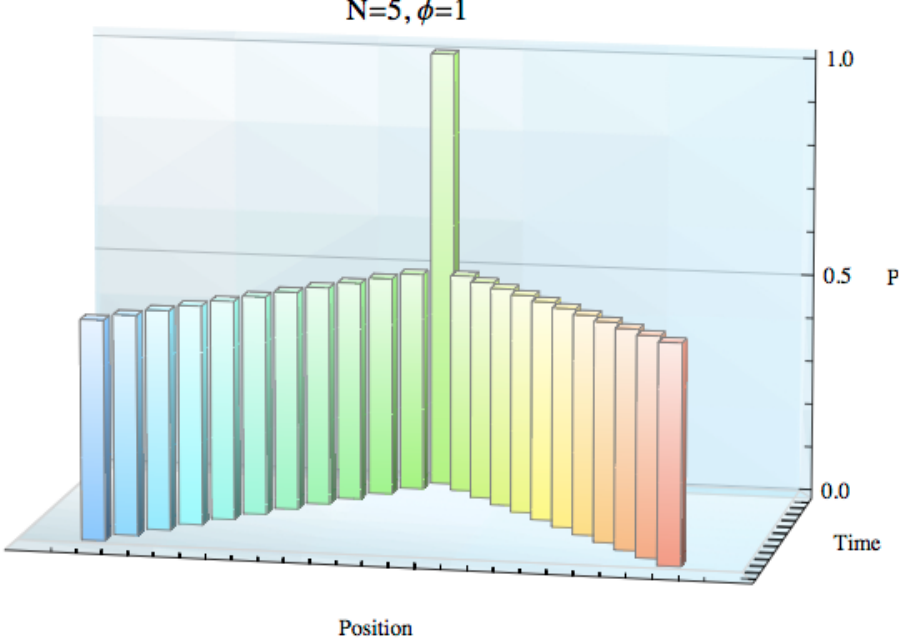}
\caption{Time evolution over 12 steps for a wise old man walk with $N=5$ and memory function from Eq. \ref{eq:enhanced_diff}. (top) \mbox{$\phi=0.5$}, (bottom) $\phi=1$. Diffusion is enhanced in both directions compared to Fig. \ref{fig:wise_phi}(left). For $\phi=1$ we observe that the walker linearly shoots off in both directions. In general this will not be the case, but comes as a result of the choice of initial state \mbox{$\ket{\psi_\mathrm{in}} = (\ket{0,-1,\dots,-1}+\ket{0,+1,\dots,+1})/\sqrt{2}$}, yielding the maximum possible dispersion.} \label{fig:enhanced_diff}.
\end{figure}

\begin{figure}
\includegraphics[width=\columnwidth]{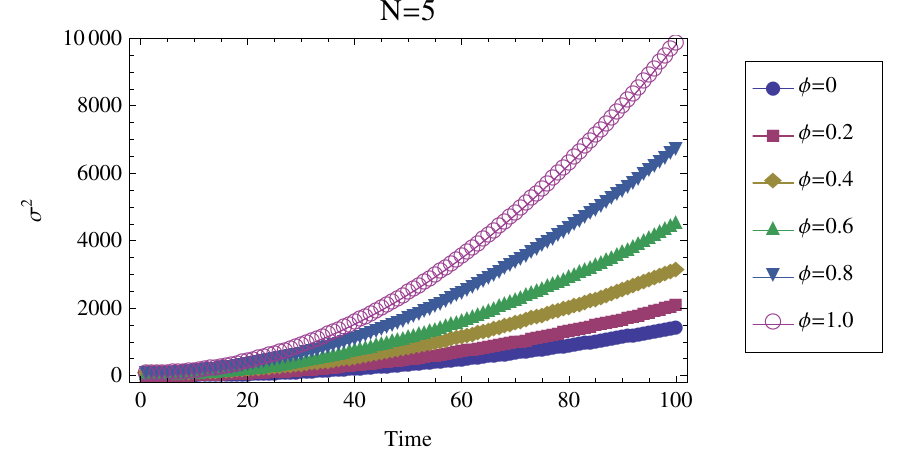}
\caption{(Colour online) Evolution of the variance of the probability distribution against time for an $N=5$ wise old man walk for different history parameters $\phi$ and memory function from Eq. \ref{eq:enhanced_diff}. As $\phi$ increases, diffusion increases. All curves fit closely to quadratic distributions.} \label{fig:wise_spread_enh}
\end{figure}

The two memory functions discussed previously had the effect of reducing dispersion. One can also choose memory functions which enhance diffusion. If we choose our memory function to be,
\begin{equation} \label{eq:enhanced_diff}
\theta = \frac{\pi}{4} - \phi \frac{\pi}{4} \frac{|\sum_{i=1}^{N-1}c_i|}{N-1},
\end{equation}
diffusion will be enhanced for larger $\phi$. When the history is consistently in the same direction the coin reduces to the identity, and to a Hadamard when the history is balanced. This is shown in Fig. \ref{fig:enhanced_diff} (with a symmetrized input state). In Fig. \ref{fig:wise_spread_enh} we plot the variance against time for various $\phi$ with memory function from Eq. \ref{eq:enhanced_diff}. Once again all curves fit closely to quadratic distributions.

The memory function from Eq. \ref{eq:enhanced_diff} is just Eq. \ref{eq:second_diff} with a negative value of $\phi$. Thus, for the memory function in Eq. \ref{eq:second_diff} there are three regimes. When $\phi=0$ we observe usual Hadamard ballistic spreading. When \mbox{$0<\phi\leq 1$} we observe reduced diffusion. And when \mbox{$-1\leq \phi<0$} we observe enhanced diffusion. Note that \mbox{$\sigma^2\propto t^2 \,\, \forall \, \phi$}.

\subsection{A physical model for history dependent coined walks}
\label{physmodel}
Let us consider how to physically realise a history dependent coin rotation operation.  For the memory function defined in Eq. \ref{eq:theta} we need to perform a conditional rotation on the $N$th coin based on the states of the $N-1$ coins that specify the remembered past of length $N-1$  \footnote{Physical realisations of memory functions Eqs. \ref{eq:second_diff}, \ref{eq:enhanced_diff} are more difficult to achieve and we do not discuss them here.}.  We do this by an appropriate modification of the quantum walk step: \mbox{$MSC\rightarrow MSU(\phi)C$} with the action defined:
\begin{eqnarray} \label{eq:ket_rules}
C: && \ket{x,c_1,\dots,c_N} \to \sum_j A_{c_N,j}^{(x)}(\frac{\pi}{4}) \ket{x,c_1,\dots,c_{N-1},j}, \nonumber \\
U: && \ket{x,c_1,\dots,c_N} \to \sum_j U_{c_N,j}^{(x)}(\phi) \ket{x,c_1,\dots,c_{N-1},j}, \nonumber \\
S: && \ket{x,c_1,\dots,c_N} \to \ket{x+c_N,c_1,\dots,c_N}, \nonumber \\
M: && \ket{x,c_1,\dots,c_N} \to \ket{x,c_N,c_1,\dots,c_{N-1}},
\end{eqnarray}
 where
\begin{equation}
U^{(x)}(\phi)= e^{i \frac{\pi}{4}\sigma^x_N}e^{-i \frac{\pi\phi(x)}{4(N-1)}\sum_{j=1}^{N-1}\sigma_j^z\otimes\sigma^z_{N}}e^{-i \frac{\pi}{4}\sigma^x_N} 
 \end{equation}
Here we allow for the rotation angle $\phi(x)$ to also be a function of position. Note that up to conjugation by a local rotation on the $N$th coin, this unitary is generated by an Ising-like coupling between coins \mbox{$1,\ldots ,N-1$} and the $N$th coin.   Effectively, the unitary $U(\phi)$ performs a conditional rotation on the $N$th coin with an angle that is proportional to the polarisation \mbox{$S^z=\sum_{j=1}^{N-1}\sigma^z_j$} of all the other coins in memory.

\section{Quantum walks with memory on general graphs}

In our studies so far we have restricted ourselves to regular graphs. As discussed, we can generalise to regular graphs of constant order by increasing the dimensionality of the memory states. Next we consider the question whether the formalism can be generalised to arbitrary graph structures. To address this question we turn to the formalism presented in Ref. \cite{bib:RohdeSchreiber11}, which introduced an approach for modelling multi-walker quantum walks on arbitrary graphs. We will restrict ourselves to the case of a single walker, in which case the formalism is,
\begin{eqnarray} \label{eq:coin_step_def}
&C:& \, \ket{x,c} \to \sum_{j\in n_x} A_{cj}^{(x)} \ket{x,j}, \nonumber \\
&S:& \, \ket{x,j} \to \ket{j,x}.
\end{eqnarray}
Here $c$ no longer specifies a direction, but rather a vertex. Thus, after the coin operator $c$ can be interpreted as the vertex the walker will hop to at the next step, and after the step operator as memory of the previous vertex. $n_x$ is the \emph{neighbourhood} of vertex $x$, i.e. the vertices connected to $x$ by an edge.

Next one might assume we can logically generalise this to have memory by letting our state vectors be of the form \mbox{$\ket{x,c_1,\dots,c_N}$}, as before. The problem with doing this is that the number of possible values of $c_N$ is in general larger than the size of the neighbourhood of $x$, $|n_x|$. Thus, no unitary matrix $A^{(x)}$ can be defined which maps the $c_N$ vertex memory value to a vertex in the neighbourhood of $x$. Therefore, in this intuitive choice of formalism, it is not possible to unitarily define a quantum walk with memory. If it is possible at all, a more novel approach will have to be employed, or alternately non-unitary operators could be employed, which would result in mixed states.

A different approach to considering arbitrary graphs is to instead focus on regular graphs, such as a regular 2D lattice, but allow the coin matrices $A^{(x)}$ to be different at each vertex. By this approach irregularity can be modelled, even though the underlying graph is regular. For example, to model a vertex which does nothing in a 2D lattice, the corresponding coin matrix could be defined as the identity matrix \mbox{$A^{(x)}=I=\mathrm{diag}(1,1,1,1) \,\, \forall \, x$}. This would have the effect of propagating the walker straight through the vertex without entering a superposition.

% Multiple walkers
\section{Multiple walkers}

Next we generalise our formalism to multiple walkers. To do this we adopt the \emph{walker operator} formalism presented in Ref. \cite{bib:RohdeSchreiber11}. Here we replace position/coin state vectors with creation operators (in an optical context these are photon creation operators), of the form $w(x,c_1,\dots,c_N)^\dag$. Using this approach a system with multiple walkers can be easily modelled by adding additional walker operators. The transformations on the walker operators are defined analogously to Eq. \ref{eq:ket_rules},
\begin{eqnarray} \label{eq:operator_formalism}
C: && w(x,c_1,\dots,c_N)^\dag \to \sum_j A_{c_N,j}^{(x)} w(x,c_1,\dots,c_{N-1},j)^\dag, \nonumber \\
S: && w(x,c_1,\dots,c_N)^\dag \to w(x+c_N,c_1,\dots,c_N)^\dag, \nonumber \\
M: && w(x,c_1,\dots,c_N)^\dag \to w(x,c_N,c_1,\dots,c_{N-1})^\dag.
\end{eqnarray}

Using this formalism a single walker state would be defined as \mbox{$w(x,c_1,\dots,c_N)^\dag\ket{0}$}, where $\ket{0}$ represents an empty graph (the vacuum state in an optical context). With $n$ walkers this generalises to \mbox{$w(x^{(1)},c_1^{(1)},\dots,c_N^{(1)})^\dag \dots w(x^{(n)},c_1^{(n)},\dots,c_N^{(n)})^\dag\ket{0}$} (up to normalisation), where $c_i^{(j)}$ is the coin associated with memory element $i$ of walker $j$.

We will not present any numerical results for multiple walkers here, since the complexity of the system grows exponentially with the number of walkers, making it computationally intractable for even a modest amount of memory. This will be discussed further in Sec. \ref{sec:classical_sim}.

Bosons only interfere with one another when they are indistinguishable. Thus, in a continuous-time quantum walk, where no coins are present, multiple walkers will only interfere when they share the same position. In an ordinary single-coined discrete-time quantum walk, multiple walkers will only interfere when they share the same position \emph{and} coin values. Similarly, multiple walkers with memory will only interfere when they share the same position \emph{and} their entire memory history is the same. For this reason it is to be expected that there will be much less quantum interference taking place between walkers with memory, since the probability of different trajectories sharing the same memory history drops exponentially with the size of the memory.

% Memory as a decoherence model
\section{Memory as a decoherence model}

We have seen that full memory reduces our quantum walk to a walk with classical statistics. We now develop some further intuition as to why this is the case.

The most general form of discrete-time evolution is the quantum process formalism -- this formalism captures unitary evolution, measurement, as a well as decoherence processes. It is well known that any quantum process can be expressed as a primary system $P$ interacting with an environment system $E$, where the environment is traced out and only the primary system is observed \cite{bib:NielsenChuang00}. Thus, any quantum process can be expressed as, 
\begin{equation} \label{eq:quant_proc}
\rho = \mathcal{E}(\ket\psi\bra\psi) = \mathrm{tr}_E(U_{P,E}\ket\psi\bra\psi U_{P,E}^\dag),
\end{equation}
where $U_{P,E}$ is a unitary acting jointly on the primary and environment systems, which in general entangles the two systems. When $U_{P,E}$ is an entangling operation, the quantum process will in general yield a mixed state.

\begin{figure}[!htb]
\includegraphics[width=\columnwidth]{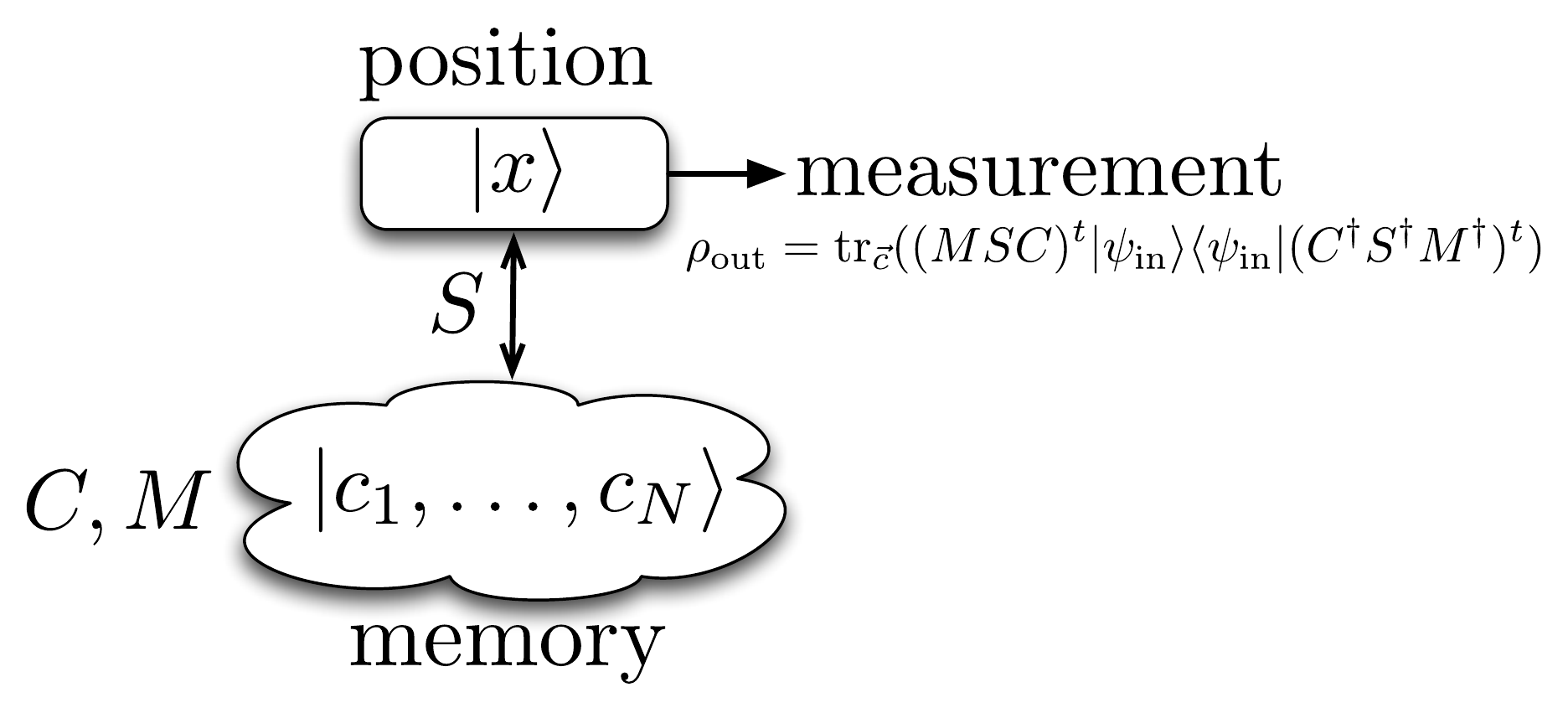}
\caption{Visualising memory as an environment system. The coin and memory update operators act only on the memory subsystem, while the step operator couples the memory environment with the position. At the end of the evolution we trace out the environment, only observing the position subsystem, yielding, in general, a mixed state.} \label{fig:decoherence_model}
\end{figure}

In the case of our quantum walk with memory this unitary takes the form \mbox{$U_{P,E} = MSC$}. Here $C$ and $M$ act only on the coin subsystem, whereas $S$ is an entangling operation that couples the two subsystems. Thus, $U_{P,E}$ is entangling. Following evolution we only measure $P$, tracing out $E$, a model analogous to Eq. \ref{eq:quant_proc}. This is illustrated in Fig. \ref{fig:decoherence_model}.

To understand how full memory reduces a quantum walk to have classical statistics, let us consider the first step in the evolution of a quantum walk. Since we are just considering the first step, it suffices to introduce just one coin, in which case the evolution over one step is given by,
\begin{eqnarray}
\ket{\psi_\mathrm{in}} &=& \ket{0,+1}, \nonumber \\
C \ket{\psi_\mathrm{in}} &=& (\ket{0,+1} - i\ket{0,-1})/\sqrt{2}, \nonumber \\
SC \ket{\psi_\mathrm{in}} &=& (\ket{1,+1} - i\ket{-1,-1})/\sqrt{2}.
\end{eqnarray}
After $SC$ we have a maximally entangled state across two positions. Tracing out the coin system we obtain the maximally mixed state across two positions,
\begin{equation}
\mathrm{tr}_{\vec c}(SC \ket{\psi_\mathrm{in}} \bra{\psi_\mathrm{in}} C^\dag S^\dag) = (\ket{1}\bra{1} + \ket{-1}\bra{-1})/2,
\end{equation}
what one would obtain for a single step of a classical random walk. Obviously if we repeat this many times, each time employing a fresh coin, our evolution will proceed classically, since introducing a fresh coin, maximally entangling it with the position and then tracing out the coin is simply a balanced classical coin-flip.

Thus, full memory, which is equivalent to introducing a new coin at each step, yields a classical walk, provided the coins are all traced out prior to measurement of the position. On the other hand, with partial memory we expect to see purely classical statistics for \mbox{$t\leq N$}, after which quantum behaviour emerges since there are quantum correlations between past and present coin values. With no memory (i.e. a single coin) we are re-coupling to the same environmental degree of freedom at each step, and thus we do not reduce to a classical walk since each coin-flip is not independently random.

From a non-rigorous, intuitive point of view, this result is perhaps not surprising. As we increase the size of the environment we are effectively making the joint system more macroscopic, and increased decoherence is intuitively not surprising.

Based on the arguments presented, we expect that decoherence is a generic feature of quantum systems with memory where the memory is not measured.

% Walks with Spatial Randomness
\section{Walks with Spatial Randomness}
 \begin{figure}[!htb]
\includegraphics[width=\columnwidth]{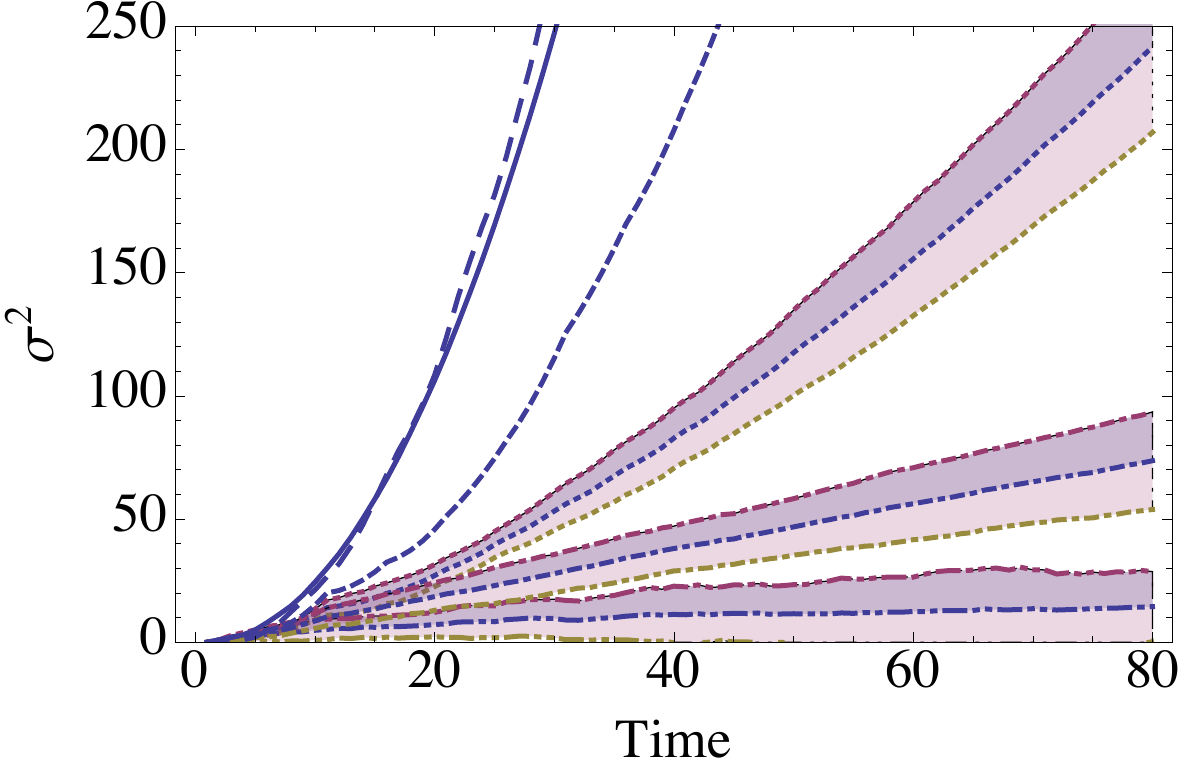}
\caption{(Color online) Comparison of various quantum walk behaviours.  Quantum walks on a line with:  five recycled coins, memory function from Eq. \ref{eq:theta}, no randomness (long dashed); one coin and no randomness, i.e. the standard quantum walk (solid); five recycled coins, trivial memory function, and no randomness (short dashed); five recycled coins, memory function Eq. \ref{eq:theta}, spatially random coin flip (dotted); five recycled coins, trivial memory function, spatially random coin flip (dot-dashed); one coin, spatially random coin flip (dot-dot-dashed).  The plots with randomness were generated by averaging over $30$ realisations of random spatially dependent angles for the coin flip, where the mean is plotted in the middle and one standard deviation plotted above and below.} 
\label{Randomfig}
\end{figure}

It is known that spatial randomness in various forms can give rise to localization of quantum walkers.  In particular, using a spatially dependent coin with rotation angles that are chosen randomly and quenched, leads to a localized distribution.  Localization is a phenomenon that arises due to coherence, specifically destructive interference between amplitudes for propagation away from the origin.  Joye and Merkli \cite{bib:JoyeMerkli} and Ahlbrecht \emph{et al.} \cite{bib:Ahlbrecht1} show that dynamical localisation occurs for a spatially inhomogeneous coin, when the coin parameters are chosen randomly from continuous or discrete sets.  If temporal randomness is also present, e.g. when the coin parameters change in time, then localisation does not occur and instead the quantum walk looks classically diffusive \cite{bib:Ahlbrecht2}. In contrast, the walker spreads ballistically or diffusively, but no localisation has been observed.  

 We examine how the presence of a memory affects such behaviour.   To do so we use the memory function in Eq. \ref{eq:theta} but with an angle that varies randomly in space:
\begin{equation}
\theta = \frac{\pi}{4} + \phi(x)\frac{\pi}{4}\frac{\sum_{i=1}^{N-1}c_i}{N-1}
\end{equation}
where $\phi(x)$ is chosen randomly in the range \mbox{$-1\leq \phi(x) \leq 1$}, independently at each position.

Results for various types of walks are shown in Fig. \ref{Randomfig}.  The three walk scenarios without spatial randomness all behave quantum mechanically, i.e. they have quadratic dispersion.  The walker with $5$ coins and a memory function spreads notably faster than that without, and indeed after a crossover time of $t=15$ spreads faster than the standard quantum walk with one coin.  We cannot confirm that this is universal behaviour, however for various initial conditions the walk with memory function does spread quadratically, and it was already proven that the walk with finite number of spins (less than $t$) and trivial memory function also spread quadratically \cite{bib:Brun03}. The results with spatial randomness are more striking. The one coin walk with randomness shows localisation as predicted. However the $5$ coin walk with trivial memory function does not localise and indeed has a variance scaling linearly with $t$.  We interpret this as due to the fact that the multiple coins provide a sufficiently large environment such that the coherence necessary to localise is lost. This is in contrast to the case without randomness where enough coherence is preserved to maintain quadratic dispersion as discussed above. In effect this is indicating that the coherence necessary to observe localisation is more sensitive to decoherence than is the coherence to obtain quantum mechanical speedup. Even more surprising is that the quantum walk with memory function spreads much faster in the presence of randomness than either of the other two cases. For this size simulation we are unable to confirm the asymptotic behaviour of this case though the linear coefficient of the variance does dominate for \mbox{$t<80$}. We conjecture the sub-ballistic behaviour of the multi-coined walks in the presence of randomness as a consequence of the coins introducing an effective temporal randomness to the dynamics by reintroducing past spatial randomness at each new step of the walk. The memory function we have used tends to average out the effect of spatial randomness from earlier steps which could explain why it is spreading faster than the case of the walk with trivial memory function.
 
 % Experimental construction
\section{Experimental construction}

Linear optics quantum computing (LOQC) \cite{bib:KLM01, bib:Kok05, bib:KokLoevett10} has become one of the more promising candidates for implementing large-scale quantum information processing devices. We now briefly discuss the prospects for linear optics implementation of quantum walks with memory.

\begin{figure}[!htb]
\includegraphics[width=\columnwidth]{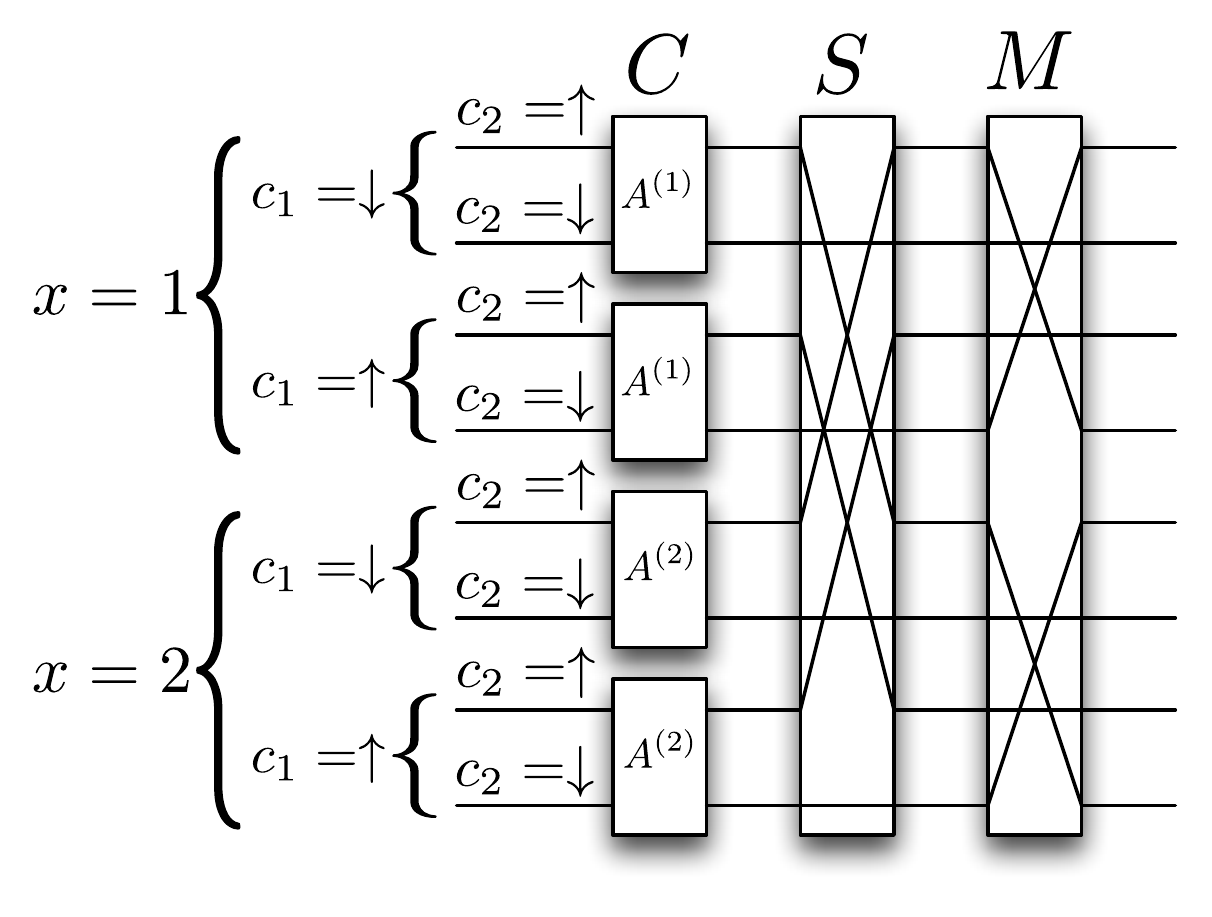}
\caption{Explicit linear optics construction of an $N=2$ quantum walk with two position states. The `wires' represent optical modes and are bundled together to represent different configurations of position and coin values. The coin operators act on $c_2$ only, while $S$ and $M$ are permutation operators.} \label{fig:lo_const}
\end{figure}

When using single photons to represent walkers, this formalism can always be experimentally constructed using just passive linear optics networks (i.e. beamsplitters and phase-shifters) in a manner similar to that described in Ref. \cite{bib:RohdeFedrizzi12}. As a simple example, the explicit linear optics construction of an $N=2$ quantum walk with two positions is depicted in Fig. \ref{fig:lo_const}. Each `wire' represents an optical mode, and wires are bundled together to represent the different position/coin combinations. In Fig. \ref{fig:lo_const} the coin operator acts only on the $c_2$ space, while $S$ and $M$ are permutation operators as per Eq. \ref{eq:operator_formalism}.

In general, with an $N$-element memory, $d$ position states and $|c|$-dimensional coins, there will be \mbox{$d\cdot |c|^N$} wires. Thus the number of modes required for experimental implementation grows exponentially with the length of the memory, making large-memory demonstrations unviable.

Experimental construction of wise of man walks would be particularly challenging, as it would require implementing conditional coin operations, coherently controlled by some arbitrary function of the memory history. This is largely beyond the scope of present-day linear optics implementations.

Nonetheless, linear optics demonstrations of walks with small memories and without history-dependence are viable in the near future. An experiment like that shown in Fig. \ref{fig:lo_const} could be readily constructed with present-day technology.

% Classical simulation & computational complexity
\section{Classical simulation \& computational complexity} \label{sec:classical_sim}

As discussed earlier, the number of optical modes required to implement a quantum walk with memory grows as \mbox{$d\cdot |c|^N$}. This makes experimental construction challenging, since the number of optical modes grows exponentially with the size of the memory. With $n$ walkers, the number of unique basis states grows as \mbox{$O((d\cdot |c|^N)^n)$}. Thus, the complexity of brute-force classical simulation of this quantum system grows exponentially against both the number of walkers and the length of the memory. Therefore classical simulation of such systems will be limited to walks with a modest number of walkers and small memories, and for this reason we do not present numerical results for multi-walker scenarios. Results for multiple walkers \emph{without} memory were previously presented in Ref. \cite{bib:RohdeSchreiber11}.

From an algorithmic perspective, we are interested in systems with exponential complexity, since otherwise the system will be classically simulatable and exponential quantum speedup will obviously not be possible. It was noted in Ref. \cite{bib:RohdeSav10} that if the size of the graph is `efficient' (i.e. \mbox{$d\cdot |c|^N = O(\mathrm{poly}(p))$}, where $p$ is the size of the problem), then the only way to achieve exponential complexity is with the addition of multiple walkers, as noted above.

Because quantum walks with memory can be implemented with just linear optics, then, with multiple walkers, such systems are a subset of the Boson-sampling problem described by Aaronson \& Arkhipov \cite{bib:AaronsonArkhipov10}, who presented strong evidence that such systems are classically hard to simulate. It was shown in Ref. \cite{bib:RohdeFedrizzi12} that an ordinary single-coined quantum walk with multiple walkers is already universal for Boson-sampling (complexity class denoted \textbf{BosonSampP}). Thus, at best, adding memory to the walk will also be universal for Boson-sampling -- adding memory cannot extend the complexity class of the system beyond \textbf{BosonSampP}. It is known that \mbox{$\mathbf{BosonSampP} \subseteq \mathbf{SampBQP}$} (the sampling version of \textbf{BQP}), and it is strongly believed that \mbox{$\mathbf{BosonSampP} \subset \mathbf{SampBQP}$} \cite{bib:AaronsonArkhipov10}. Therefore we expect multi-walker quantum walks with memory will present no algorithmic advantage over multi-walker single-coined quantum walks, although they are nonetheless likely to be classically hard to simulate, and quantum walks with memory (on efficiently sized graphs) are likely not universal for quantum computation. However, on exponentially large graphs it is known that both discrete- and continuous-time quantum walks \emph{without} memory are \textbf{BQP} \cite{bib:Childs09, bib:Lovett10}, and we don't rule out that the same applies to quantum walks with memory.
 
% Conclusion
\section{Conclusion}

We have presented a formalism for quantum walks where the walker has arbitrary amounts of memory of its previous history. 
%We demonstrated that sufficient memory reduces the output statistics to classical statistics, even though no decoherence is introduced. Thus, goldfish are more quantum than elephants. As the amount of memory increases, we observe a gradual transition from quadratic spread (quantum statistics) to linear spread (classical statistics). 
We examined the effects of history-dependent coins and introduced the memory function, which manipulates the diffusive properties of the walk as a function of the walker's history. With appropriate choice of memory function, diffusion can be reduced, enhanced, or be made symmetric or asymmetric. We also studied the effects of measurement of the memory registers on the evolution of the walk. Measurement of the oldest memory elements has the effect of `resetting' the quantum walk to begin at a new position that is a function of the measurement outcomes. In the case of a two-dimensional walk on a regular rectangular lattice, memory eliminates spatial entanglement between the two spatial degrees of freedom, yielding a separable spatial distribution, even if entangling coins are employed. 

A key observation from our numerics arises in the context of spatial randomness.  We found that in settings where single coined walks localize, multi-coin walks do not.  Moreover, walks with non-trivial memory function appear to spread faster than those which only recycle coins.  We conjecture that this occurs due to the the memory function averaging out the mechanism which induces an effective temporal randomness to the walk due to recycling coins.  It has been shown that for one dimensional walks with both spatial and temporal randomness, the former dominates to lead to classical diffusion \cite{bib:Ahlbrecht3}.  Whether walks with memory can counteract this effect is unknown.

Finally, we explicitly demonstrated how to physically construct an optical quantum walk with memory. However the required physical resources grow exponentially with the length of the memory. Thus, experimental construction of a quantum walk with large memories is limited. We presented a formalism for multi-walker quantum walks with memory and discussed the prospects for classical simulation. Unfortunately the required classical resources grow exponentially against both the number of walkers and the size of the memory. Therefore classical simulation is challenging beyond a few walkers with even modest memory. We argued that a multi-walker quantum walk with memory has no algorithmic advantage over its single-coined counterpart. Algorithmic complexity aside, quantum walks with memory may be interesting to pursue nonetheless owing to their unique physical properties, such as their rich diffusive characteristics, which may be interesting for quantum simulation applications.

% Acknowledgments
\begin{acknowledgments}
This research was conducted by the Australian Research Council Centre of Excellence for Engineered Quantum Systems (Project number CE110001013). We thank Klaus Rohde and Scott Aaronson for helpful discussions.
\end{acknowledgments}

% Bibliography
\bibliography{paper}

\end{document}